\documentclass[doublecol]{epl2} 
\usepackage{amsmath}
\usepackage{amssymb}
\usepackage{graphicx}
\usepackage{color}

\title{Synchronization transitions in globally coupled rotors
in presence of noise and inertia: Exact results}
\shorttitle{Synchronization transitions in globally coupled rotors}

\author{Maxim Komarov\inst{1,2} \and Shamik Gupta\inst{3} \and Arkady
Pikovsky\inst{1,2}}
\shortauthor{M. Komarov \etal}

\institute{                    
  \inst{1} Department of Control Theory, Nizhni Novgorod
University, Gagarin Av. 23, 606950, Nizhni Novgorod, Russia\\
  \inst{2} Department of Physics and Astronomy, Potsdam University,
Karl-Liebknecht-Str 24, D-14476, Potsdam, Germany \\
  \inst{3}  Laboratoire de Physique Th\'{e}orique et Mod\`{e}les
Statistiques (CNRS UMR 8626), Universit\'{e} Paris-Sud, Orsay, France
}
\pacs{05.45.Xt}{Synchronization; coupled oscillators}
\pacs{05.70.Fh}{Phase transitions: general studies}
\pacs{05.70.Ln}{Nonequilibrium and irreversible thermodynamics}

\abstract{We study a generic model of globally coupled rotors that
includes the effects of noise, phase shift in the coupling, and
distributions of moments of inertia and natural frequencies of
oscillation. As particular cases, the setup includes previously studied
Sakaguchi-Kuramoto, Hamiltonian and Brownian mean-field, and
Tanaka-Lichtenberg-Oishi and Acebr\'on-Bonilla-Spigler models. We derive an exact solution of the
self-consistent equations for the order parameter in the stationary
state, valid for arbitrary
parameters in the dynamics, and demonstrate nontrivial phase transitions to synchrony that include reentrant synchronous regimes.
}
\begin{document}
\def\be{\begin{equation}}
\def\ee{\end{equation}}
\def\bea{\begin{eqnarray}}
\def\eea{\end{eqnarray}}
\def\fr{\frac}
\def\l{\label}
\def\th{\theta}
\newcommand{\dd}{\mbox{d}}
\maketitle

\section{Introduction}
Synchronization in a large population of coupled oscillators of
distributed natural frequencies is a
remarkable example of a nonequilibrium phase transition. The paradigmatic
minimal model to
study synchronization is the one due to Kuramoto, introduced
almost 40 years ago~\cite{Kuramoto-75}, based on an earlier work by Winfree~\cite{Winfree-67}. 
Over the years, many details of
the Kuramoto model~\cite{Kuramoto-84,Acebron-etal-05}, and applications to
various physical~\cite{Wiesenfeld-etal-96},
chemical~\cite{Aldridge-etal-76},
biological~\cite{Michaels-etal-87}, engineering~\cite{Dorfler-etal-13}, and even social 
problems~\cite{Neda-etal-03} have been addressed in the literature.

The Kuramoto model comprises oscillators that are described by their
phases, have natural frequencies given by a common distribution, 
and are subject to a global mean-field coupling. The phases follow a
first-order dynamics in time. In the simplest setup of a purely
sinusoidal coupling without a phase shift, and for a unimodal distribution
of frequencies, the model exhibits a continuous
(second-order) 
transition from an unsynchronized to a synchronized phase as the
coupling constant exceeds a critical threshold. The phase
transition appears as a Hopf bifurcation for the complex order parameter.

The dynamics of the Kuramoto model is intrinsically dissipative. When all the 
oscillators have the same  frequency, the analogue of the model in the realm of 
energy-conserving Hamiltonian dynamics is the so-called Hamiltonian mean-field 
model (HMF)~\cite{Inagaki-93,Antoni-Ruffo-95}. In this case, the dynamical
equations are the Hamilton equations: the oscillator phases follow a
second-order dynamics in time, i.e. the units are in fact not oscillators, but rotors. 
In order to include the effects of interaction with 
an external
heat bath, it is natural to consider the HMF evolution in presence of a Gaussian 
thermostat. In the resulting
Brownian mean-field (BMF) model, the dynamical equations are damped and 
noise-driven \cite{Chavanis-11,Chavanis-13}.
Both the HMF and the BMF model have an equilibrium stationary state that
exhibits a continuous phase transition between a synchronized phase at low
values of energy/temperature and an unsynchronized phase at high values.
On considering the BMF model with non-identical oscillator frequencies,
the dynamics violates detailed balance leading to a nonequilibrium stationary state
(NESS) \cite{Gupta-Campa-Ruffo-14}. In the overdamped limit, the
dynamics reduces to that of the noisy Kuramoto model involving Kuramoto
dynamics in presence of Gaussian noise, which was introduced to model
stochastic fluctuations of the natural frequencies in time~\cite{Sakaguchi-88}. The resulting
phase diagram is complex, with both continuous and first-order 
transitions~\cite{Gupta-Campa-Ruffo-14}.

In this work, we study a generic model of globally coupled rotors, 
in which two types of deviations from 
equilibrium are included: (i) distribution of torques acting on the
rotors, similar to the distribution of frequencies in the Kuramoto model,
and (ii) a phase shift
in the coupling, that makes the latter 
non-Hamiltonian. We consider the rotors to have quite generally
different moments of inertia given by a common
distribution~\cite{Restrepo-etal-14}. Our setup
includes as special cases previously studied
Sakaguchi-Kuramoto~\cite{Sakaguchi-Kuramoto-86}, Hamiltonian and
Brownian mean-field,
Tanaka-Lichtenberg-Oishi~\cite{Tanaka-Lichtenberg-Oishi-97}, and
Acebr\'on-Bonilla-Spigler~\cite{Acebron-etal-98,Acebron_etal-00} models.  

The basic roadblock in studying out-of-equilibrium dynamics, in
particular, in characterizing the resulting long-time NESSs is the lack of a 
framework that allows to treat such states on a general footing, akin to
the one for equilibrium steady states {\em \`{a} la} Gibbs-Boltzmann. Even for simple nonequilibrium models, obtaining the steady
state distribution has been a {\em tour de
force}~\cite{Derrida-etal-1993}, while in many
cases, the analytical characterization of the steady state has so far
been elusive, thereby requiring one to resort to numerical simulations and
approximation methods~\cite{Privman-05}.

In this backdrop, it is remarkable that for our system of study, we are
able to characterize exactly the NESS under quite general
conditions. In the thermodynamic limit $N \to \infty$, we study the system by analyzing the
Kramers equation for the evolution of the single-rotor phase space
distribution. Using the combination of an analytical approach to solve the Kramers equation in
the steady state~\cite{Risken-89} and a novel self-consistency
approach~\cite{Komarov-Pikovsky-13a,Omelchenko-Wolfrum-12}, we
formulate an exact equation for the complex order parameter as a function
of the relevant parameters of the system, for arbitrary distributions of torques and moments
of inertia. As applications of our approach, we provide for suitable
and representative choices of the distribution functions several
nontrivial illustrations of transitions to synchrony, including in some cases interesting reentrant synchronous regimes.

\section{Basic model}
The equations of motion for the $i$-th rotor read
\be
m_i\ddot{\phi_i}+\gamma
\dot{\phi_i}=\gamma\nu_i+\mathcal{K}R\sin(\psi-\phi_i-\beta)+\sqrt{\gamma
T}~\eta_i,
\l{eq:eom-1}
\ee
where dots denote differentiation with respect to time $\tau$. Here, $\phi_i$ 
is the angle of the
$i$-th rotor with moment of inertia $m_i$, $\gamma$ is the friction
constant, $\mathcal{K}$ is the coupling constant, $\beta$ is the phase shift
parameter, $T$ is the temperature in units of the Boltzmann constant, 
while $R$ and $\psi$ constitute the complex order parameter of the problem:
$R(\tau)e^{i\psi(\tau)} \equiv \sum_{j=1}^N e^{i\phi_j(\tau)}/N$.
Note that $R$ is the magnitude while $\psi$ is the phase of the mean
field acting on the rotors. In the dynamics (\ref{eq:eom-1}),
the term $\gamma \nu_i$ is the external torque acting on the
$i$-th rotor.
The parameters of the rotors, namely, their frequencies $\nu_i$'s and the moments
$m_i$'s, are quenched random variables sampled from a common
distribution $g(\nu,m)$. 
The noise $\eta_i(\tau)$ is Gaussian, with $\langle \eta_i(\tau) \rangle=0, ~~\langle \eta_i(\tau)\eta_j(\tau')
\rangle=2\delta_{ij}\delta(\tau-\tau')$. 
To reduce the number of relevant parameters, we introduce dimensionless variables $t \equiv \tau T/\gamma$, $M_i \equiv
m_iT/\gamma^2$, $K=\mathcal{K}/T$, $\omega_i=\gamma \nu_i/T$, in terms of
which Eq. (\ref{eq:eom-1}) becomes (with dot now denoting derivative with respect
to $t$)
\be
M_i\ddot{\phi_i}+\dot{\phi_i}=\omega_i+KR\sin(\psi-\phi_i-\beta)+\eta_i,
\l{eq:eom-2}
\ee
involving two dimensionless parameters $K$ and $\beta$; here, $\eta_i(t)$ is a
Gaussian white noise with $\langle \eta_i(t) \rangle=0,\langle
\eta_i(t)\eta_j(t')=2\delta_{ij}\delta(t-t')$.
Additional parameters describe the distribution $G(\omega,M)$ of dimensionless
natural frequencies and moments.

For the dynamics (\ref{eq:eom-2}), we seek for NESS with non-zero,
uniformly-rotating mean field, which generally has a frequency $\Omega$ different from
the mean frequency of the natural frequencies distribution.
Transforming to the reference frame rotating with frequency
$\Omega$, as $\psi \equiv \Omega t+\psi_0, \phi_i \equiv \Omega t+
\theta_i+\psi_0-\beta$,
where $\psi_0$ is 
a constant, Eq. (\ref{eq:eom-2}) reads
\be
M_i\ddot{\th}_i+
\dot{\th}_i= \omega_i-\Omega-KR\sin \th_i+\eta_i(t),
\l{eq:eom-3}
\ee
with $R(t)e^{i\beta} \equiv \sum_{j=1}^N e^{i\th_j(t)}/N$.
From now on, we focus on analyzing the dynamics of (\ref{eq:eom-3}).

At this point, it is instructive to link model (\ref{eq:eom-3})
to previously studied setups.
In the absence of a distribution of frequencies and moments, and for $\beta=0$, 
Eq. (\ref{eq:eom-3}) describes the BMF model. This model has an equilibrium
stationary state in which the system exhibits a continuous phase transition 
between a synchronized ($R \ne 0$) and
an unsynchronized
($R =0$) phase at the critical coupling $K_c=2$ \cite{Chavanis-11,Chavanis-13}. 
Presence of frequencies in the dynamics
(\ref{eq:eom-3}) drives
the system to a NESS
\cite{Gupta-Campa-Ruffo-14}. For $M=0$, the dynamics (\ref{eq:eom-3}) is that of the 
Sakaguchi-Kuramoto
model with the inclusion of noise, which shows a continuous synchronization phase 
transition across $K_c(D)$ described for $\beta=0$ in Ref.~\cite{Sakaguchi-88}.
In our normalization, the intensity of noise is set to one, thus the
noiseless situation corresponds to the limit $K,D\to\infty$. 
For this noiseless dynamics, the case of rotors with the same moments of
inertia, and with $\beta=0$ defines the Tanaka-Lichtenberg-Oishi model ~\cite{Tanaka-Lichtenberg-Oishi-97}. 
In Ref.~\cite{Acebron_etal-00}, 
for model (\ref{eq:eom-2}) with $\beta=0$ and without the distribution of
moments (the Acebr\'{o}n-Bonilla-Spigler model), 
a linear stability analysis and an approximate treatment
of the transition to synchrony have been performed. In both these works, a first-order transition
to synchrony was revealed.

\section{Thermodynamic limit: The Kramers equation and its self-consistent
stationary solution}
We now consider the dynamics (\ref{eq:eom-3}) in the thermodynamic limit $N
\to \infty$. In this limit, the dynamics is 
characterized by the single-rotor conditional distribution
$\rho(\th,v,t|\omega-\Omega,M,t)$, which
gives at time $t$ and for the given set of parameters $(\omega,M)$ the 
fraction of rotors with angle $\th$ 
and angular velocity $v=\dot\th$. The distribution is $2\pi$-periodic in
$\theta$, and obeys the normalization $\int_0^{2\pi} d\th
\int_{-\infty}^{\infty} dv~\rho(\th,v,t|\omega-\Omega,M,t)=1$, while evolving 
toward a stationary distribution
following the Kramers (Fokker-Planck) equation
\cite{Gupta-Campa-Ruffo-14}
\be
\fr{\partial \rho}{\partial t}=-v\frac{\partial \rho}{\partial
\th}+\frac{\partial}{\partial
v}\Big[\frac{1}{M}
\Big(v-\omega+\Omega+A\sin\th\Big)\rho\Big]+
\frac{1}{M^2}\frac{\partial^2 \rho}{\partial v^2},
\l{eq:Kramers}
\ee
where we have defined $A \equiv KR$. 
In the steady state, $R$ is time-independent, with 
\be
Re^{i\beta}=\int dW
G(\omega+\Omega,M)e^{i\th}\rho(\th,v|\omega,M)\equiv F(\Omega,A),
\l{eq:R-steadystate}
\ee
where $dW=d\th dv d\omega dM$. The stationary distribution $\rho(\th,v|\omega-\Omega,M)$
depends on the unknown quantities $A$ and $\Omega$, which we from now on
consider as given parameters. The representation (\ref{eq:R-steadystate})
gives the solution of the problem in a parametric form: the order parameter $R$
and the coupling parameters, $K,\beta$, are expressed as explicit
functions of $A$ and $\Omega$, as
\be
R=|F(\Omega,A)|\;,\;\; K=\frac{A}{|F(\Omega,A)|}\;,\;\;\beta={\rm arg}F(\Omega,A).
\l{eq:F-defn}
\ee
By varying $\Omega$ and $A$, while keeping the parameters of the 
distribution $G(\omega,M)$ 
fixed, we may find the order parameter $R$ as a function
of $K$ and $\beta$ (cf.~\cite{Omelchenko-Wolfrum-12,Komarov-Pikovsky-13a}). 

The stationary solution of the Kramers equation (\ref{eq:Kramers}) 
is described in~\cite{Risken-89}.
One looks for 
 a solution in the form of a double expansion in Fourier modes in $\theta$ and Hermite 
functions in $v$ as
\be
\rho =(2\pi)^{-1/2}
\Phi_0(v)\sum_{n=0}^\infty \sum_{k=-\infty}^{\infty}a_{n,k}e^{ik\theta}\Phi_n(v),
\l{eq:distr}
\ee
where $\Phi_n(v)$ are the Hermite functions:
$\Phi_n(v)=\sqrt{\alpha}/(\sqrt{2^n n!\sqrt{\pi}})
\exp[-v^2\alpha^2/2]H_n(\alpha v); ~~\alpha = \sqrt{M/2}$.
By inserting expansion (\ref{eq:distr}) into the Kramers equation
(\ref{eq:Kramers}), 
one obtains a linear system of equations for coefficients $a_{n,k}$ which can 
be solved using matrix continued fraction method \cite{Risken-89}. 
Substituting expansion (\ref{eq:distr}) into Eq. (\ref{eq:R-steadystate}), we find that  
\be
F(\Omega,A) = \sqrt{2\pi}\int d\omega\,dM G(\omega,M) a^*_{(1,0)}(\omega-\Omega,
M),
\l{eq:funF}
\ee
where $*$ denotes complex conjugation. According to the matrix continued fraction method \cite{Risken-89}, the
coefficient $a_{1,0}$ can be found from the matrix $\mathbf{H}$ as
$$
a_{(1,0)}(\nu,M)= H^{(1,0)}(\nu.M)/(\sqrt{2\pi}H^{(0,0)}(\nu.M))
$$ 
where $\mathbf{H}(\nu,M)$ is given by the following recurrent formula:
\bea
\label{eq:H}
\mathbf{H}=-\frac{1}{\sqrt{M}}\widetilde{\mathbf{D}}^{-1}\bigg(\mathbf{I}-M
\mathbf{D}\Big[\mathbf{I}-\frac{M}{2}\mathbf{D}\big(\mathbf{I}\nonumber
\\
-\frac{M}{3}\{\mathbf{D}[\mathbf{I}-\ldots]\}^{-1}\widetilde{\mathbf{D}}\big)^{-1}\widetilde{\mathbf{D}}\Big]^{-1}\widetilde{\mathbf{D}}\bigg), 
\eea
with the matrix $\mathbf{D}=ik\delta_{n,k}$, while the matrix
$\widetilde{\mathbf{D}}$ is 
\be
\label{eq:D}
\widetilde{D}^{n,k}\equiv\left(\Big(ik-\nu\Big)\delta_{n,k} -
i\Big(\delta_{n,k+1}-\delta_{n,k-1}\Big)\frac{A}{2} \right).
\ee

Combining Eqs. (\ref{eq:F-defn}), (\ref{eq:funF}), (\ref{eq:H}), (\ref{eq:D}), 
one obtains an exact equation for the complex order parameter as a function
of the relevant parameters of the system for arbitrary distributions of torques and moments
of inertia. In the following sections, we present several applications of our
approach to compute the steady state $R$ for representative choices of
the frequency and moment distribution, and parameters of the dynamics,
and highlight possible synchronization transitions.

\section{Phase transitions in the case of symmetric coupling function
(i.e., $\beta=0$)}
\begin{figure}
\centering
(a)\includegraphics[width=0.8\columnwidth]{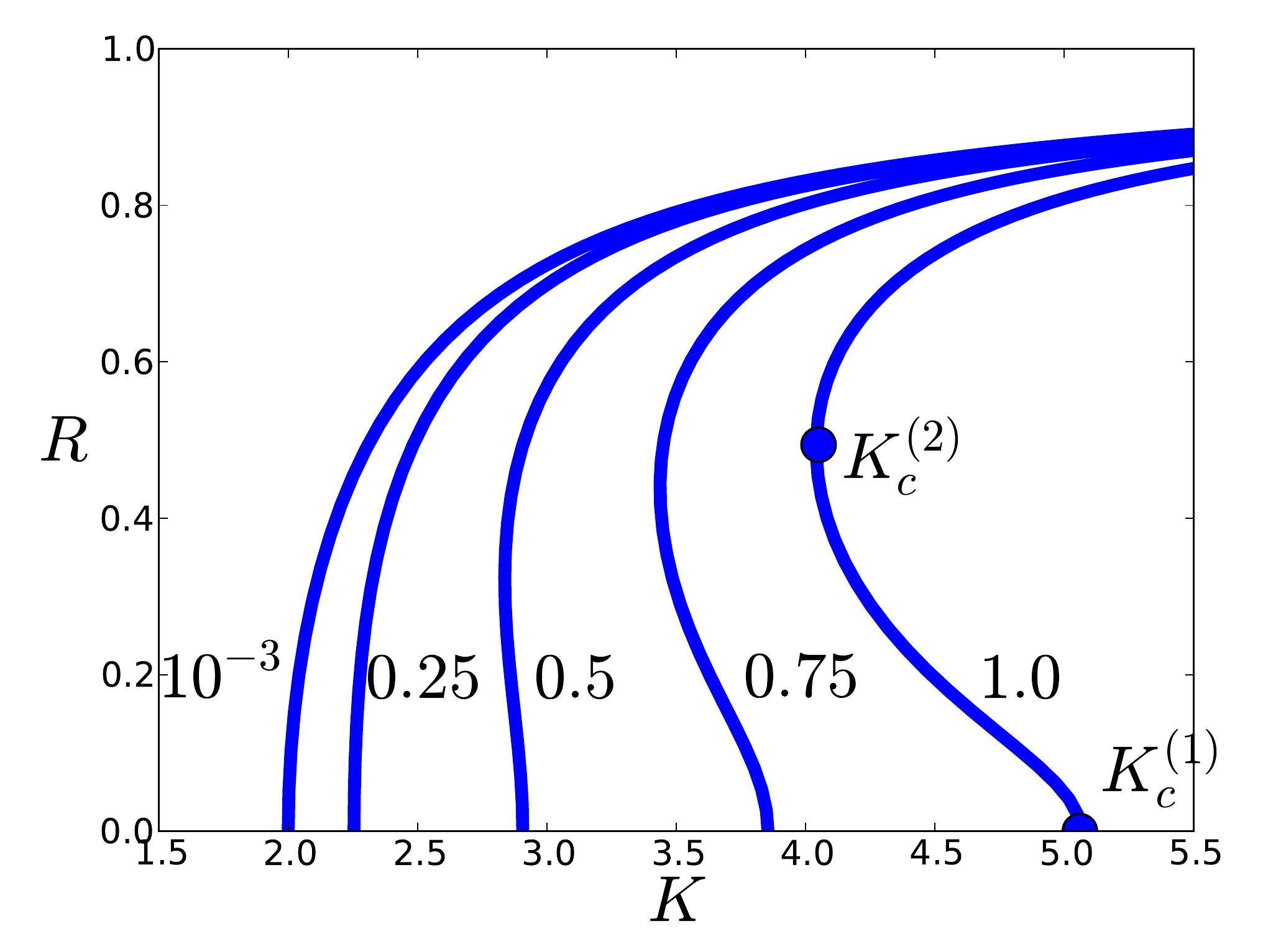}\\
(b)\includegraphics[width=0.8\columnwidth]{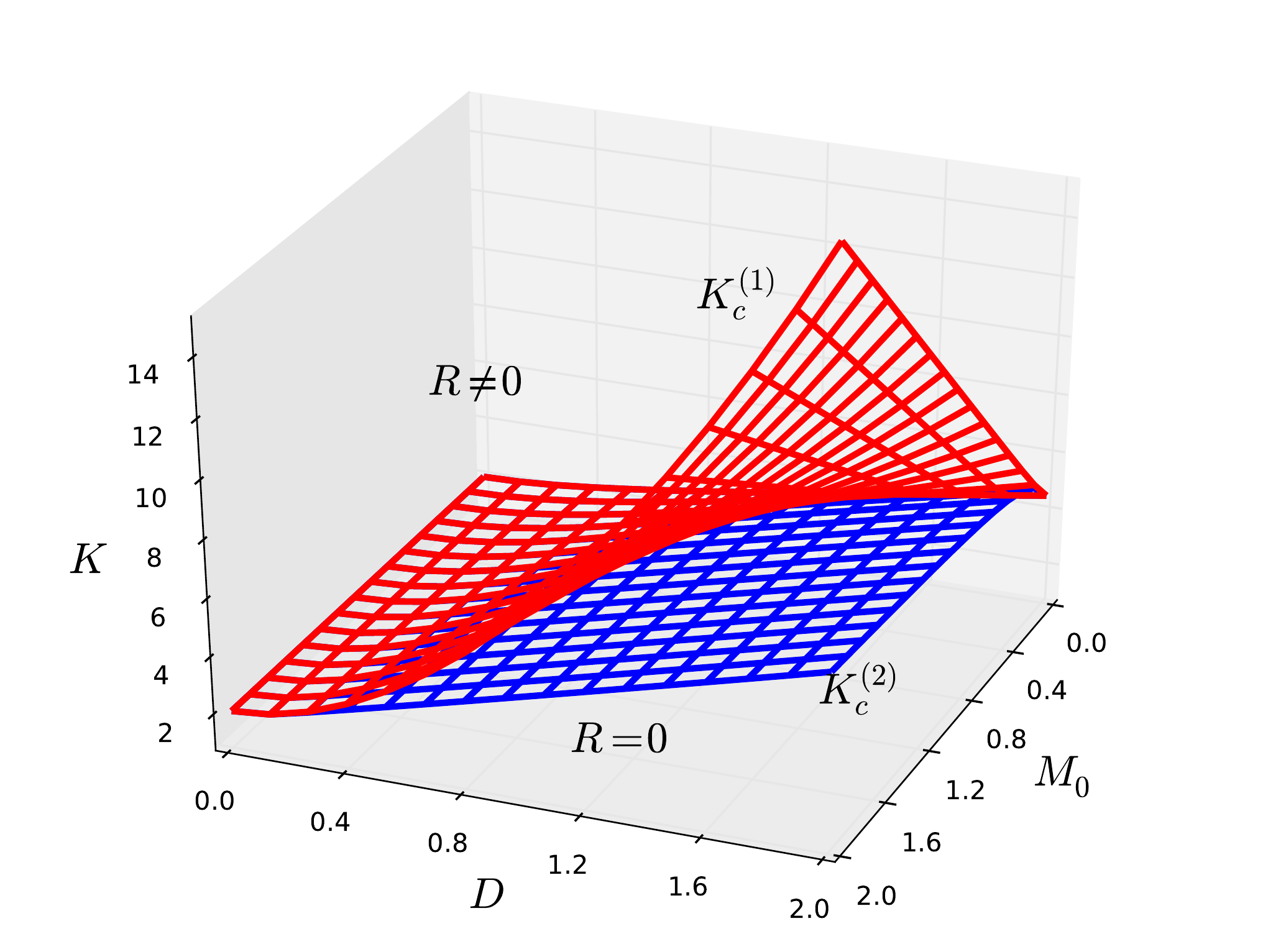}\\
(c)\includegraphics[width=0.8\columnwidth]{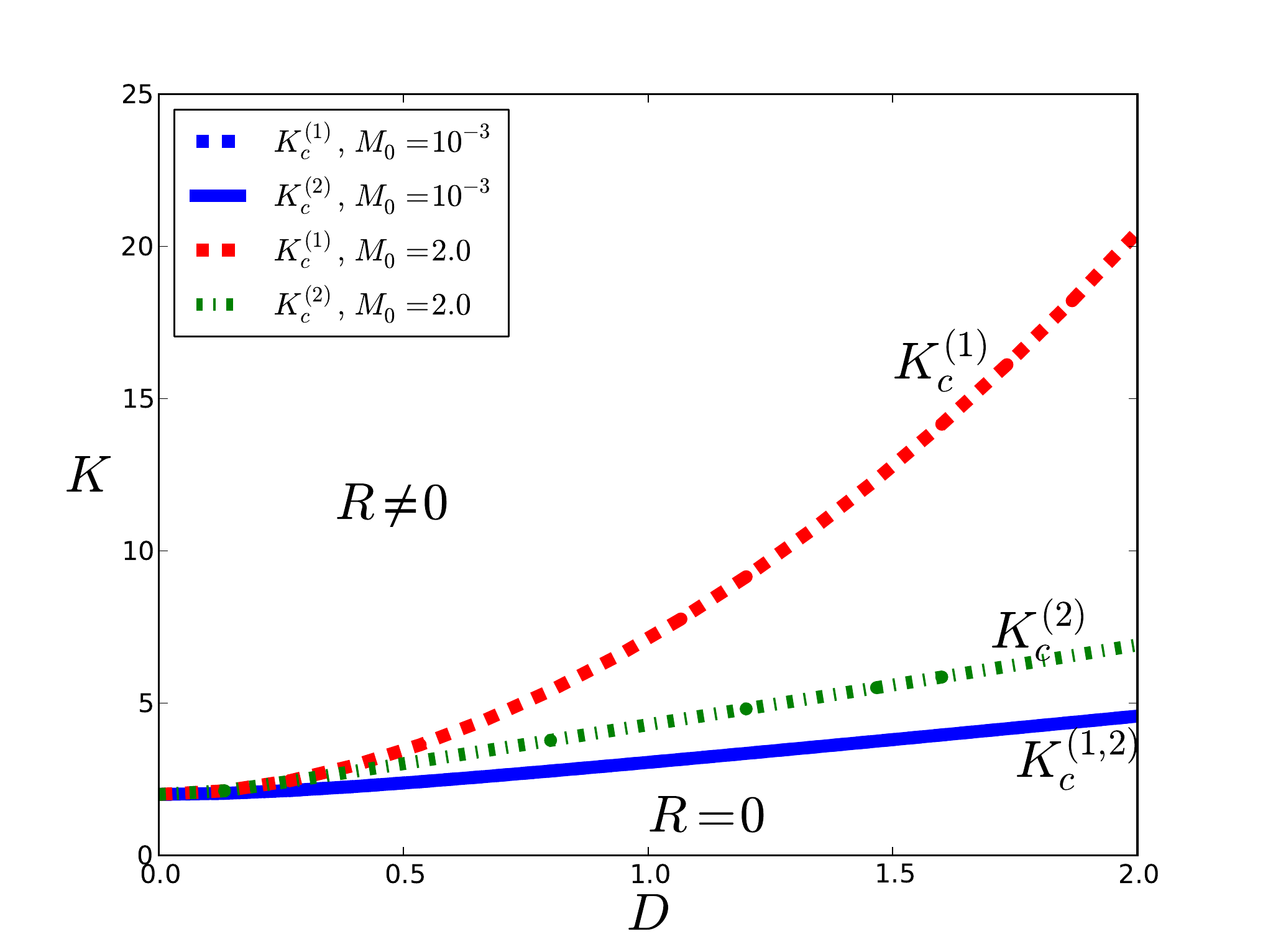}
\caption{For $G(\omega,M)=\delta(M-M_0)g(\omega)$, with $g(\omega)$
being a Gaussian with zero mean and width $D$, (a) shows the order parameter 
$R$ as a function of $K$ for $M_0=1$ and for
various values of $D$ marked in the figure. For sufficiently large $D$,
synchronized phase arises as a first-order phase transition from the
unsynchronized phase, with two characteristic thresholds, $K_c^{(1)}$ and
$K_c^{(2)}$. (b) Surfaces $K_c^{(1)}$ and $K_c^{(2)}$ in the
three-dimensional $M_0-K-D$ space. (c) Sections of the surfaces
$K_c^{(1)}$ and $K_c^{(2)}$ in (b) for different values of $M_0$.}
\l{fig:fig1}
\end{figure}

Here, we consider symmetric, Hamiltonian coupling, i.e., with the phase
shift $\beta=0$. As one can see from (\ref{eq:F-defn}), the function
$F$, given by Eqs. (\ref{eq:R-steadystate}) and (\ref{eq:funF}), should
be real. Let us choose for each
moment $M$ the distribution $G(\omega,M)$ to be symmetric about its
mean $\Omega_0$: $G(\omega-\Omega_0,M)=G(\Omega_0-\omega,M)$. Then, using Eq. (\ref{eq:funF}) and the fact that $F$ is
real, one arrives at the consistent conclusions that $\Omega=\Omega_0$
and $a_{(1,0)}(\nu,M)=a_{(1,0)}^*(-\nu,M)$. 
(For a general asymmetric distribution, like Eq.~(\ref{eq:dismw}) below, 
to be consistent with $\beta=0$, one has to vary $\Omega$ to find the value at which ${\rm arg}F(\Omega,A)=0$).
As the simplest example of such a situation, we consider the case of
equal moments, and  a Gaussian distribution with
mean zero and width $D$ for
the frequencies: $G(\omega,M)=\delta(M-M_0)g(\omega)$, with $g(\omega)=1/(\sqrt{2\pi
D^2})\exp[-\omega^2/(2D^2)]$.  In Fig.~\ref{fig:fig1}, we report the phase
diagram in the three-dimensional space of parameters $M_0,K,D$,
stressing on the synchronized and unsynchronized states. The previously known limits 
are (1) for $M=0$, when one has the line $K_c(D)$ for continuous transition 
given by $K_c(D)=2[\int_{-\infty}^\infty
d\omega~g(\omega)/(\omega^2+1)]^{-1}$ \cite{Sakaguchi-88}, (2) for
$D=0$, when one has the continuous transition point of the BMF
model given by $K_c \equiv K_c(0)=2$, independent of $M$
\cite{Chavanis-11,Chavanis-13}. 
For $M,D\ne 0$, one
expects a first-order phase synchronization transition characterized by
two thresholds $K_c^{(1)}$ and $K_c^{(2)}$, where the former is the
stability threshold of the unsynchronized phase, while $K_c^{(2)}$ is the
point at which two branches of synchronized solutions arise
\cite{Gupta-Campa-Ruffo-14}.
These observations are borne out by our numerical results depicted in
Fig. \ref{fig:fig1}. Panel (a) shows $R$ as a function of $K$ at a
fixed $M$ and for different values of $D$: one may observe that at large $D$, the
synchronized phase arises as a first-order transition from the
unsynchronized phase.

\section{Non-universal phase transitions in the case of phase shift in coupling}
In this section, we illustrate several examples of non-trivial and
non-universal phase transitions to synchrony in the case of nonzero
$\beta$, by choosing $G(\omega,M)=\delta(M-M_0)g(\omega)$, where we choose a nontrivial symmetric
$g(\omega)$ that is known to yield a non-universal transition in the
Sakaguchi-Kuramoto model~\cite{Omelchenko-Wolfrum-12}:
\be
g(\omega)=
\begin{cases}
p\frac{D - \omega}{D^2}+(1-p)\frac{qD-\omega}{q^2D^2};~~ \omega\leq qD,\\
\frac{D - \omega}{D^2};~~ \omega>qD.
\end{cases}
\l{eq:distr1}
\ee
The results are shown in Fig.~\ref{eq:fig2}. Panel (a) shows two examples of a reentrant
synchronization transition: with increase of the coupling constant, synchrony first appears but 
disappears at larger coupling, beyond which there is a second threshold.
Fig.~\ref{eq:fig2}(b)
shows that this behavior depends on the value of $M_0$ in a nontrivial way: while the re-entrance is observed for small and
large $M_0$, it is absent for intermediate values. Finally, in
Fig.~\ref{eq:fig2}(c)
we illustrate how the reentrant behavior depends on noise. As the noise intensity is set to one 
in our scheme of normalization, we simultaneously varied parameters $M_0$ and $D$ according to
$M_0=\widetilde{M}_0\sigma$, $D=D_0/\sigma$, with $\widetilde{M}_0=0.1$, $D_0=5$.
The resulting 
path in the $M_0-D$ plane corresponds effectively to variation of the noise
intensity, with $\sigma\sim T$. One can see that with increase of the 
noise, the area of nontrivial
transitions marked in grey shrinks and disappears.  
The inset illustrates how the region of the
re-entrance is determined from the solution.
It shows the values of $\beta(\Omega)$
for vanishing mean field $A=0^{+}$, i.e. at the transition points. 
At $\sigma = 0.025$ (dashed curve in (c)(inset)), there is a non-monotonic dependence
of $\beta$ on $\Omega$. Thus, in the region between the local extrema, there
are three values of $\Omega$ that give the same value of $\beta$. 
These three values correspond to the different branches in each of the $R(K)$
plots in Fig.~\ref{eq:fig2}(a) at which the value of the order parameter vanishes  (i.e. the values of $K$
at which synchrony appears, disappears, and appears again). 
At relatively strong noise ($\sigma=0.1$, solid curve in (c)(inset)), there is a
monotonic dependence of $\beta$ on $\Omega$, with only one transition to synchrony.

\begin{figure}
\centering
(a)\includegraphics[width=0.8\columnwidth]{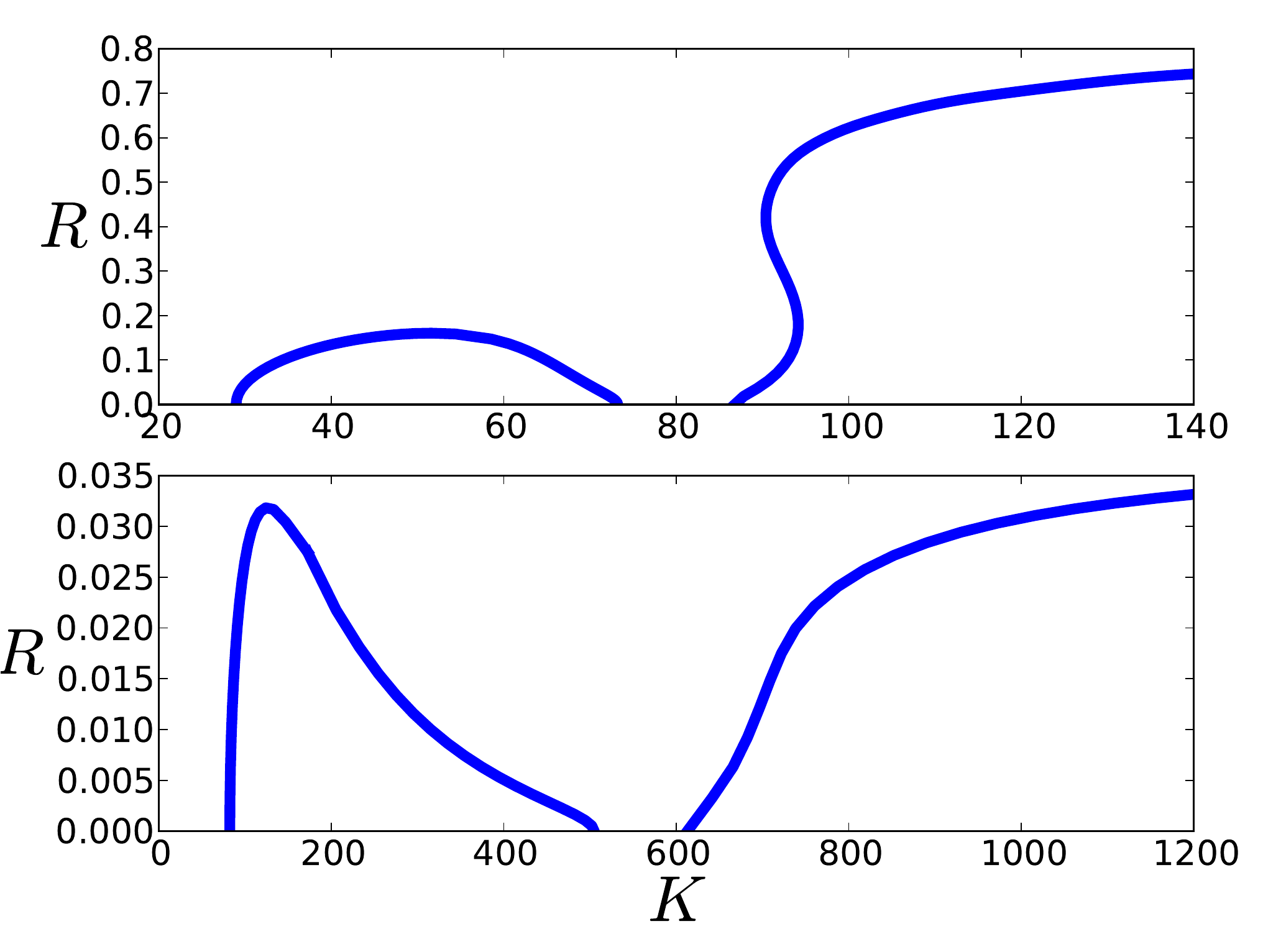}\\
(b)\includegraphics[width=0.8\columnwidth]{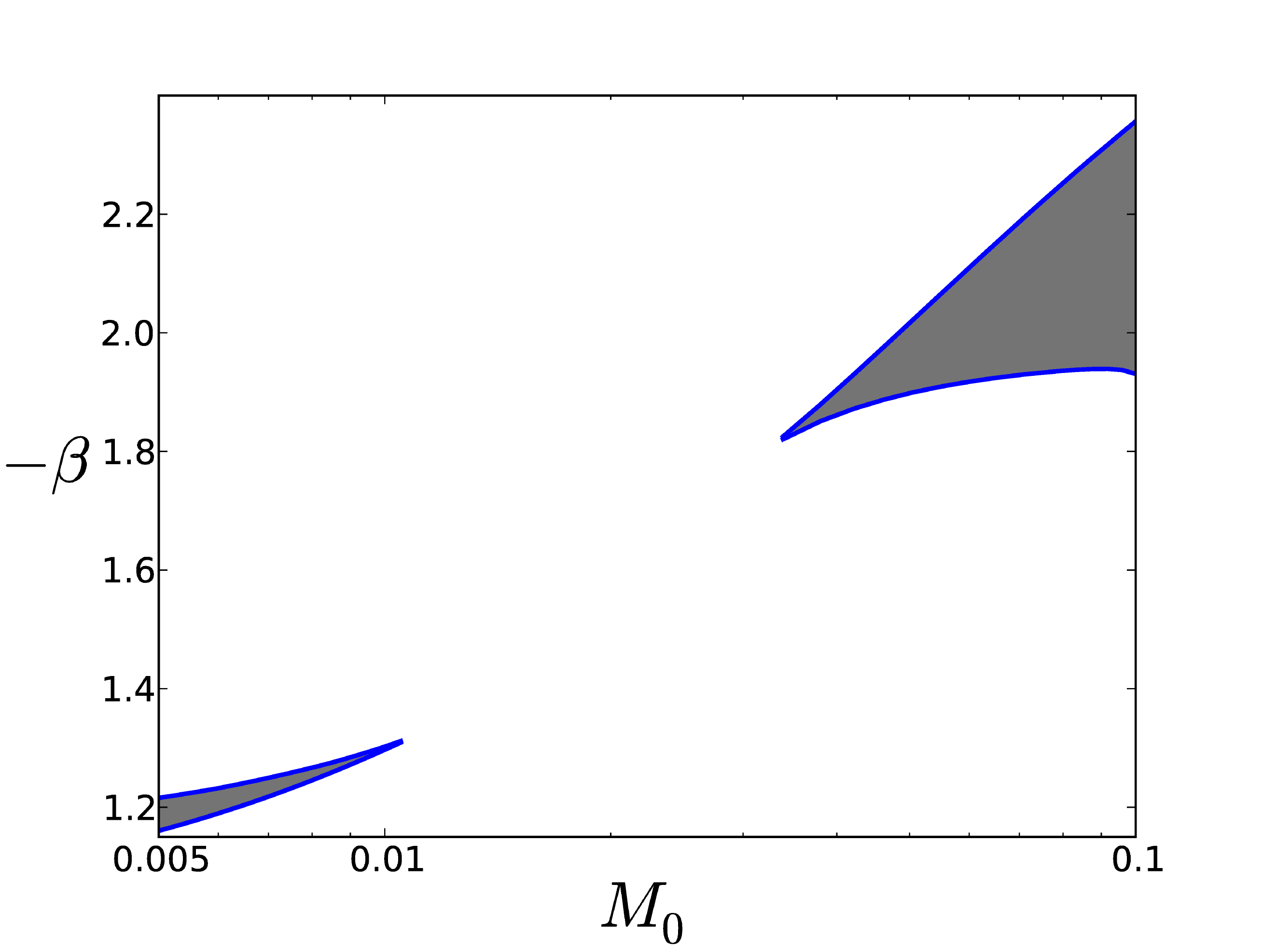}\\
(c)\includegraphics[width=0.8\columnwidth]{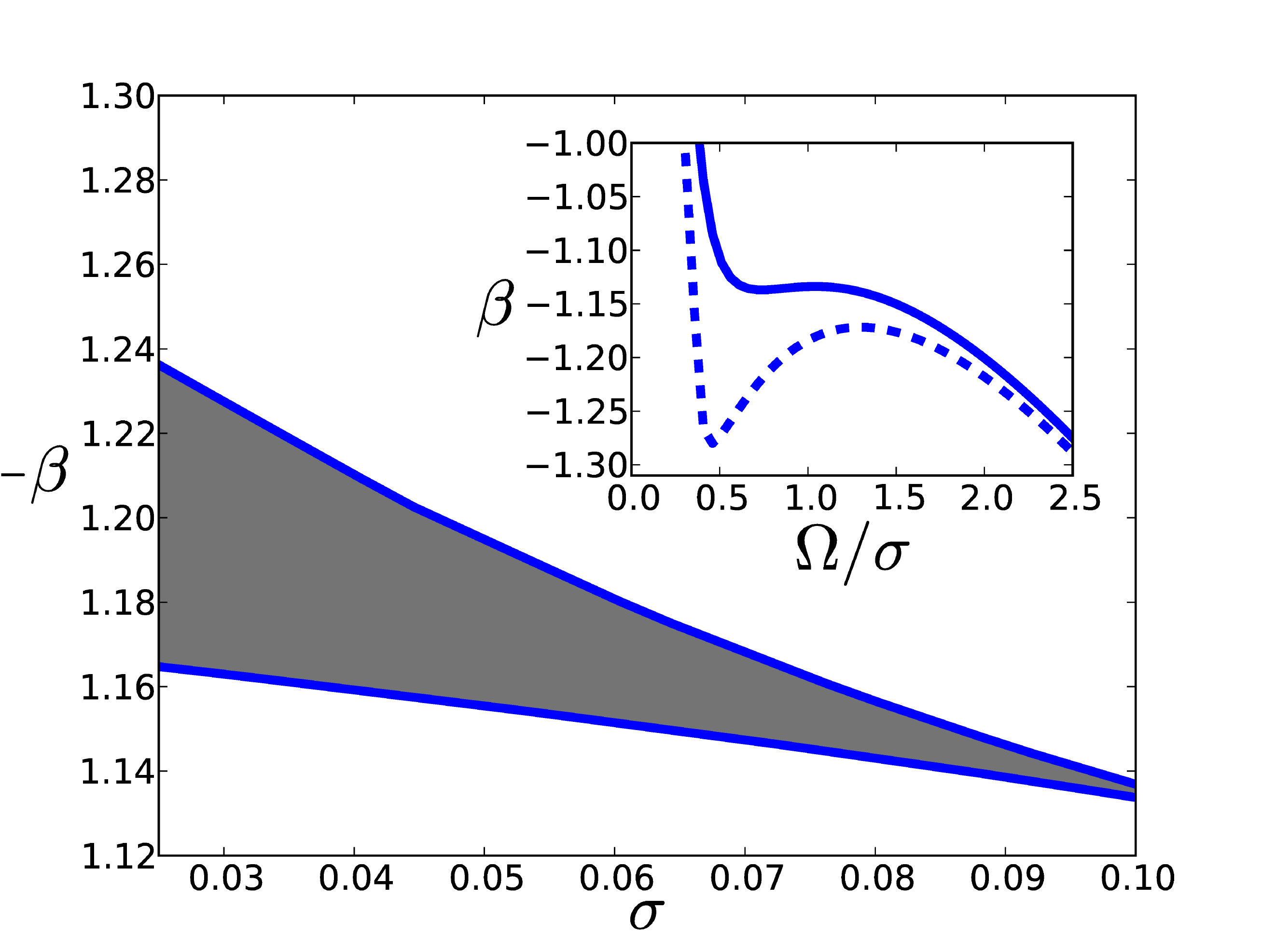} 
\caption{(a) Examples of non-universal transitions for
$G(\omega,M)=\delta(M-M_0)g(\omega)$, with $g(\omega)$ given by Eq.
(\ref{eq:distr1}). Parameters are
$M_0=5\times 10^{-3}$, $\beta = -1.164$ (upper panel), and $M_0=0.06$,
$\beta = -1.95$ (lower panel). In both cases, $p=0.6$, $D=100$, $q=0.08$. 
(b) Regions in the $(M_0,\beta)$ plane with complex non-trivial
transitions marked in grey, for the same parameters of the frequency distribution as in panel (a).
(c) Dependence of the region of re-entrance (gray regions in (b)) on the
effective noise intensity (see main text); Increase of $\sigma$ corresponds to linear 
increase of noise intensity.
}
\l{eq:fig2}
\end{figure}

\section{Phase transitions in populations with distribution of moments
of inertia}
\begin{figure}
\centering
(a)\includegraphics[width=0.8\columnwidth]{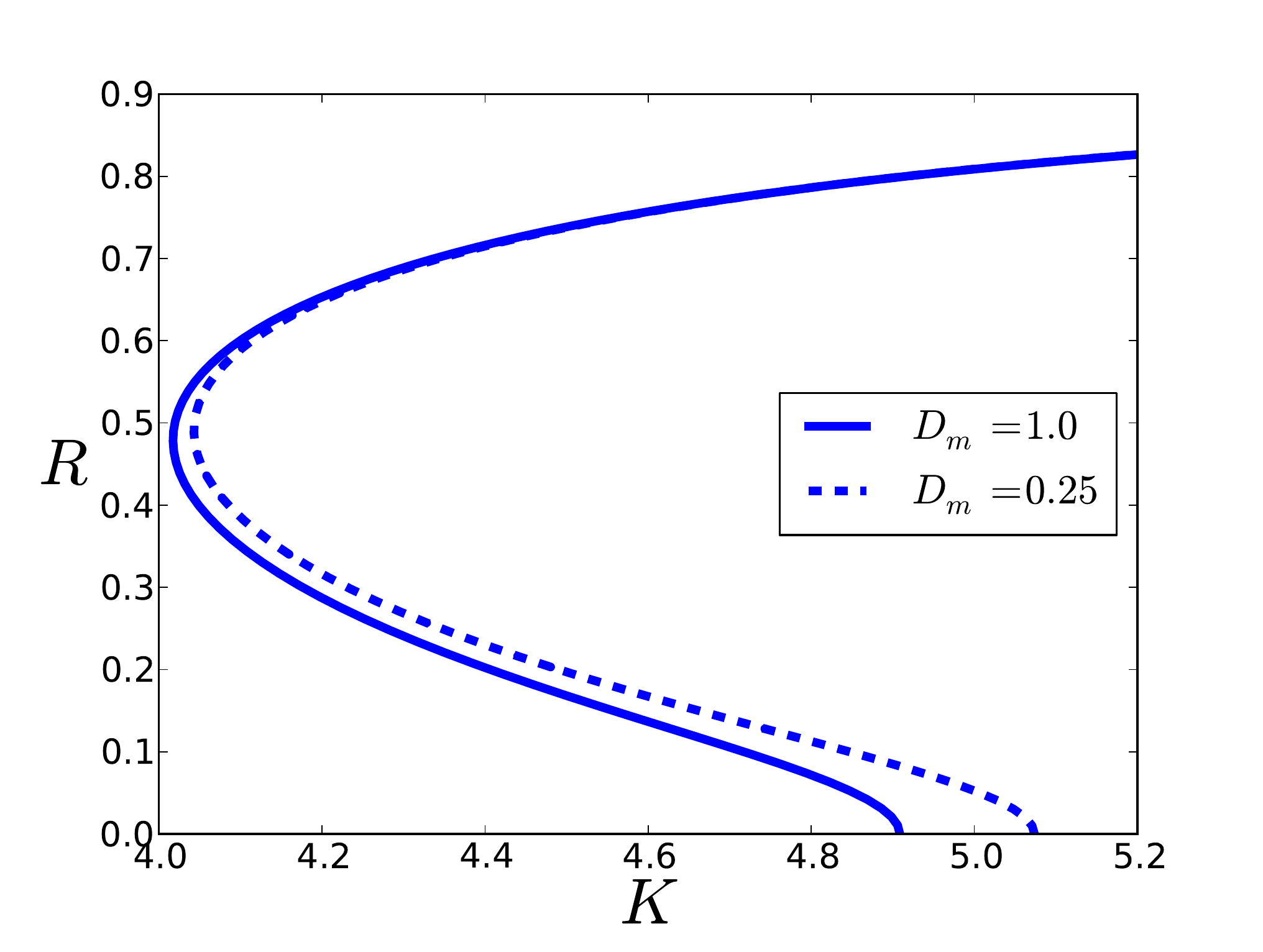}\\
(b)\includegraphics[width=0.8\columnwidth]{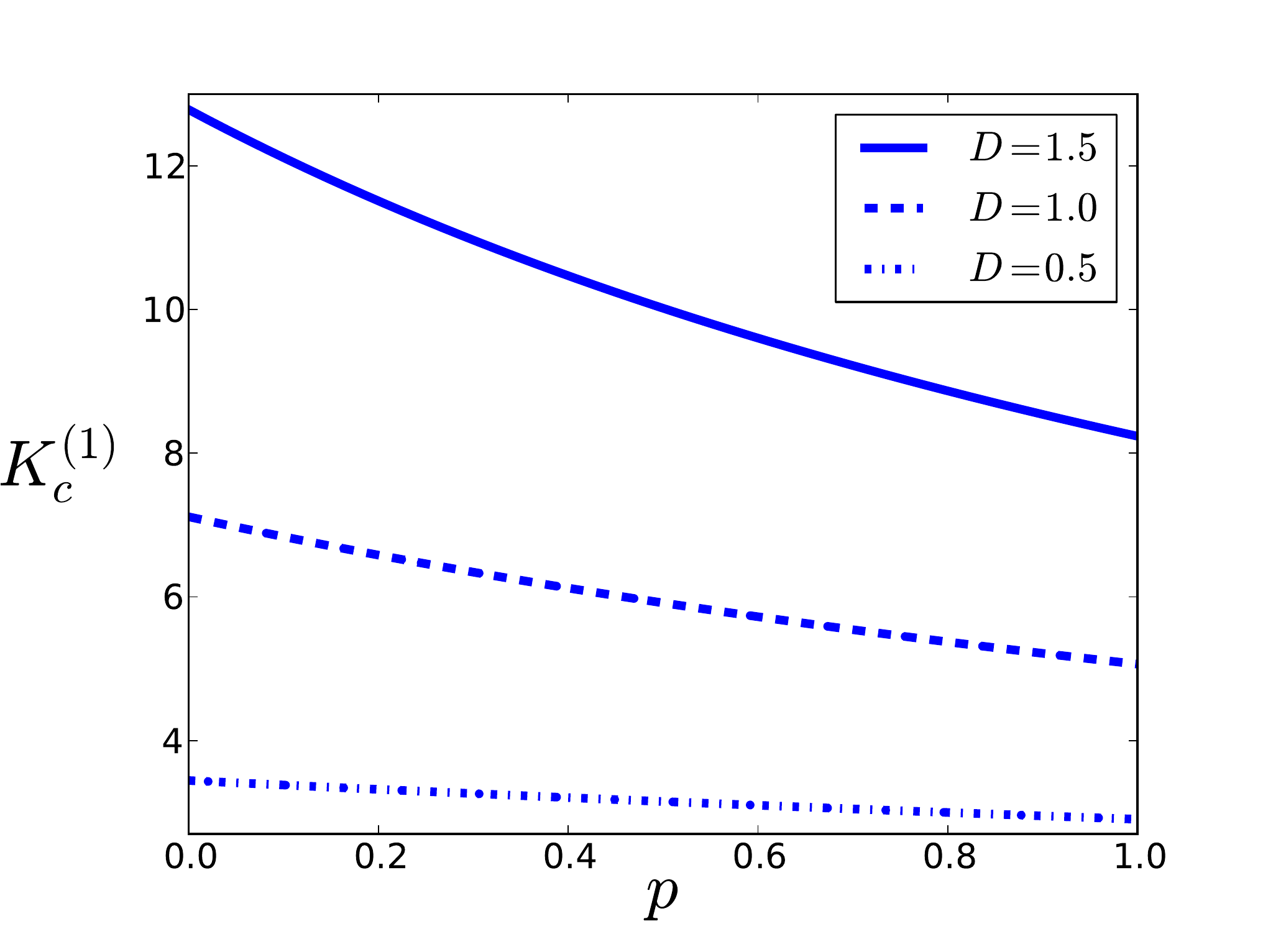}
\caption{(a) $R(K)$ for the distribution (\ref{eq:dism}) with two different 
values of $D_m$ but with the same mean moment $M_0=1.01$. 
(b) For the distribution (\ref{eq:2delta}), the figure shows the
dependence of $K_c^{(1)}$ on the parameter $p$ for different values of
$D$ (width of the frequency distribution $g(\omega)$). 
}
\l{fig:mass_tr}
\end{figure}
In this section, we present several examples of phase transitions by
choosing nontrivial distributions for both moments
and frequencies.
In the first example, we used independent distributions of moments and
frequencies: 
$G(M,\omega)=g(\omega)f(M)$, with
\be
f(M)=
\begin{cases}
\frac{1}{C}\Big[1-\fr{(M-M_0)^2}{D_m^{2}}\Big],~~ |M-M_0|\le D_m, \\
0,\quad |M-M_0|>D_m.
\end{cases}              
\l{eq:dism}
\ee
Thus, the moments are distributed according to a 
simple parabolic shape with characteristic width $D_m$ ($C$ is the
normalization constant), while for frequencies, we use a Gaussian distribution 
with mean zero and width $D$.
In Fig.~\ref{fig:mass_tr}(a), dependencies of the order parameter
on the coupling are presented for two distributions for different
$D_m$'s, but with the same mean moment $M_0$.
One can see that the more diverse is the population, the easier it is to synchronize. 
To reveal the underlying mechanism, 
we calculated the synchronization threshold $K_c^{(1)}$ in a more simple
setup of rotors having just two different moments,
i.e., the distribution is a sum of two delta functions:
\be
f(M) = p\delta(M-M_0)+(1-p)\delta(M-M_1),
\l{eq:2delta}
\ee
where we assume that $M_0<M_1$. By increasing the parameter $p$ from 0 to 1,
 we increase the fraction of light particles in the population. 
Fig.~\ref{fig:mass_tr}(b) shows that the critical coupling $K_c^{(1)}$ decreases with $p$:
One can see that addition of light particles always leads to decreasing of $K_c^{(1)}$,
implying ease of the population to synchronize with more lighter
particles; this is consistent with the result in Fig.~\ref{fig:mass_tr}(a).

In the second example we illustrate a situation, where the symmetry of the frequency distribution
is broken in a nontrivial way, through a correlation with the moments. We take 
a distribution
\be
G(\omega,M)=
\begin{cases}\frac{1}{C}\left(1-\left(\frac{\omega}{D_0}\right)^{2}\right)\\
\times \delta(M-M_0-k\omega),\;
 |\omega|\le D_0,\\
0,\quad |\omega|>D_0.
\end{cases}              
\l{eq:dismw}
\ee
where although the partial distribution of frequencies is symmetric, the overall symmetry of 
$G(\omega,M)$ is broken. In this case the frequency $\Omega$ of the 
order parameter will be non-zero even for purely symmetric coupling $\beta=0$; we 
illustrate this in Fig.~\ref{fig:asym}.

\begin{figure}
\centering
\includegraphics[width=0.8\columnwidth]{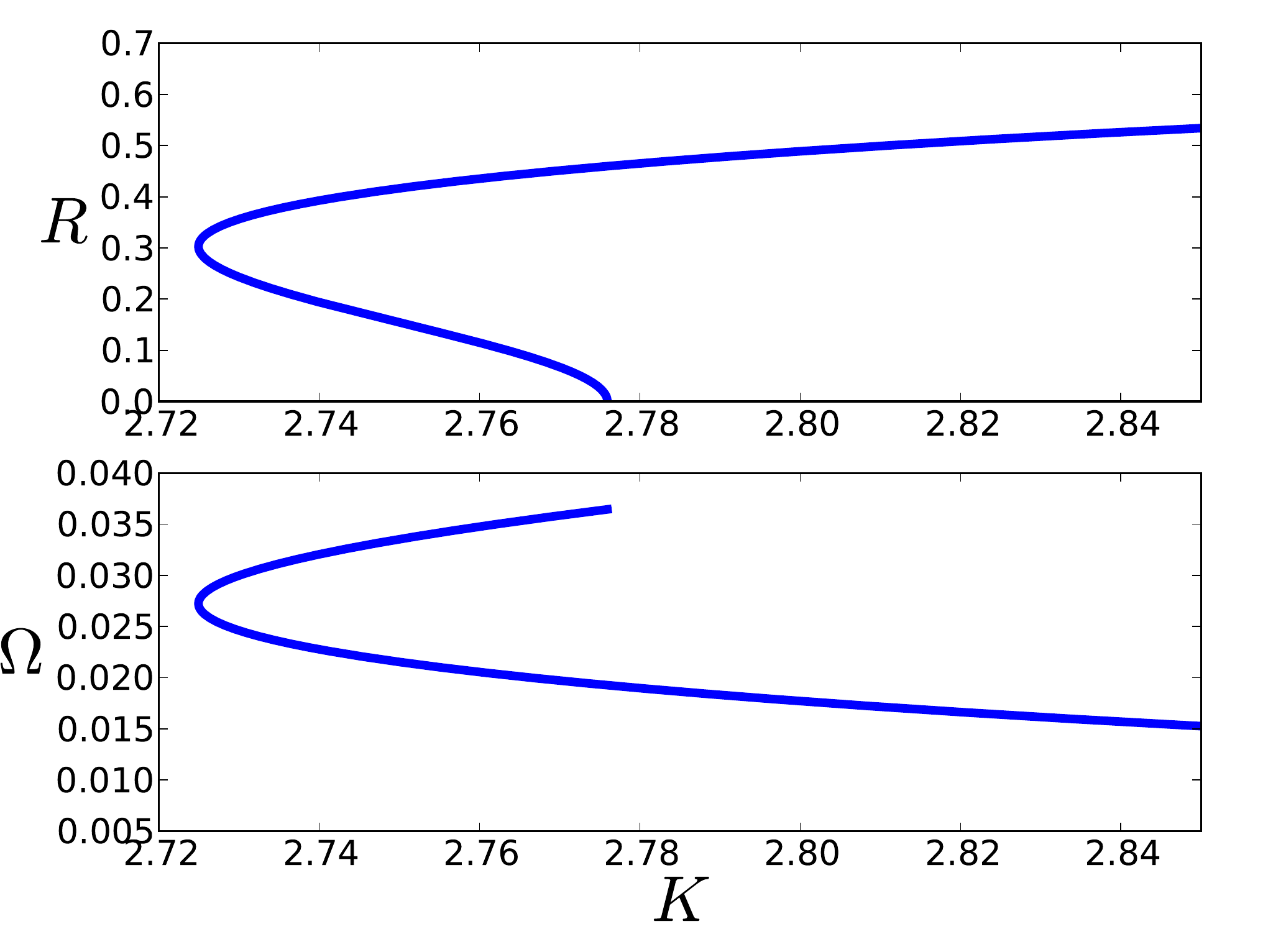}
\caption{Dependencies of $R,\Omega$ on the coupling strength $K$ for the
correlated distribution~(\ref{eq:dismw}) with $k=1$, $D_0=1$ and $M_0=1.01$. }
\l{fig:asym}
\end{figure}

\section{Conclusion}
In conclusion, we have suggested a unified analytic approach that allows to analyze
dynamics of noise-driven populations of globally coupled rotors with a phase shift in the
coupling, for arbitrary distribution of their natural
frequencies and moments. In addition to well-studied effects of inertia
that lead to a first-order transition to synchrony in the absence of a
phase shift in coupling, the method allowed us
to study more complex regimes. In the limiting case of vanishing inertia
and absence of noise, our model reduces to the Sakaguchi-Kuramoto model
of coupled phase oscillators. For the latter, reentrant transition to
synchrony~\cite{Omelchenko-Wolfrum-12}, in which two ranges of coupling
exists for observing synchrony, was observed; we demonstrated a
similar phenomenon in our model. Furthermore,
the general formulation of our model also includes populations with distributions of moments
of rotors. A nontrivial effect here is the shift of the frequency of the mean field
due to correlations between natural frequencies and the moments. 

\acknowledgments
M. K. thanks the Alexander von Humboldt Foundation for support.
S. G. acknowledges the support of the Indo-French Centre for the Promotion of Advanced
Research under Project 4604-3, the warm hospitality of the University
of Potsdam, and fruitful discussions with A. Campa and S. Ruffo.

\bibliographystyle{eplbib}
\bibliography{paper}

\end{document}